\documentclass{PoS}

\title{Jet-dominated advective systems of all mass scales}

\ShortTitle{Jet-dominated advective systems of all mass scales}

\author{\speaker{Elmar K\"ording}%
         \thanks{Marie Curie intra-European fellow}\\
        University of Southampton, UK\\
        E-mail: \email{Elmar@phys.soton.ac.uk}}

\author{Rob Fender\\
        University of Southampton, UK\\
        E-mail: \email{rpf@phys.soton.ac.uk}}

\abstract{
We show that the radio emission of black hole (BH) and neutron star (NS) X-ray binaries (XRBs) follows the analytical prediction of a jet model where the jet carries a constant fraction of the accretion power. The radio emission can therefore be used as a tracer of the accretion rate. This measure is normalised with efficiently radiating objects. As it is independent of the X-ray fluxes, the measure allows us to compare the accretion rate dependency of the bolometric X-ray luminosity of BHs and NSs. For NSs, it scales linearly with accretion rate while the scaling for BHs is quadratic - as expected for inefficient accretion flows. We find the same behaviour in AGN. This new approach uses the jet power to obtain the accretion rate. Thus, we know both the jet power and the radiated power of an accreting BH. This allows us to show that some accretion power is likely to be advected into the black hole, while the jet power dominates over the bolometric luminosity of a hard state BH.
          }

\FullConference{VI Microquasar Workshop: Microquasars and Beyond\\
                 September 18-22, 2006\\
                 Como, Italy}

\begin{document}

\section{Introduction}

The accretion rate is usually assumed to be one of the most important parameters of an accreting system. While it is possible to measure the accretion rate from the bolometric luminosity for some objects, e.g, neutron stars (NSs) as they have a stellar surface and a fairly well known accretion efficiency, it is impossible to estimate the accretion rate from the bolometric luminosity in black holes (BHs), as the accretion flow can be radiatively inefficient \cite{NarayanYi1994} and power may be advected across an event horizon or ejected in form of winds or collimated jets. Here, we show that the jet may be a better tracer of the accretion rate than the bolometric luminosity and construct an estimator of the accretion rate based on the radio luminosity.
Many of the results presented here have already been published in \cite{KoerdingFenderMigliari2006}.

During an outburst of an X-ray binary (XRB) the source can be found in several distinct states (e.g. \cite{Nowak1995,BelloniHomanCasella2005}). In quiescence and at low accretion rates the source is found in the hard state characterized by a hard power law in the X-ray spectrum. In the radio one usually observes a steady jet \cite{Fender2001}. In this state the standard accretion disk \cite{ShakuraSunyaev1973} is probably truncated and the inner part of the accretion disk turns into an inefficient accretion flow 
(e.g., an advection dominated accretion flow, ADAF \cite{NarayanYi1994,EsinMcClintockNarayan1997}). The X rays are ususally described with Comptonization models \cite{SunyaevTruemper1979,ThornePrice1975}. However, it is also possible that the compact jet may contribute to the X-ray emission or even dominate it
\cite{MarkoffFalckeFender2001,MarkoffNowakWilms2005}. When the source brightens it can stay in the hard state to fairly large accretion rates (up to 30 \% Eddington). Once the source leaves the hard state, it is often found in an intermediate state (IMS). One finds two IMSs, first, the source enters a hard IMS characterized by a hard spectral component and band-limited noise in the X-ray power spectrum. The hard IMS often shows shows strong but increasingly unstable radio emission. Additionally, a soft IMS is found, dominated by a soft spectral component with power law noise in the power spectrum (e.g.,  \cite{BelloniHomanCasella2005}). In the soft IMS no radio emission is found, but during the transition from the hard to the soft IMS an optically thin radio flare is usually observed. After leaving the IMS the source may go to the soft state, here the X-ray spectrum is dominated by a soft thermal component that is thought to originate from a standard accretion disk. Again, the radio jet is quenched in this state \cite{FenderCorbelTzioumis1999,CorbelFenderTzioumis2000} or, rather, it does not reappear. 

In the soft state a BH is well described with standard accretion theory, e.g., as expected the luminosity scales with the disk temperature as $T^4$ \cite{GierlinskiDone2004}. It is therefore likely that sources in their soft state are radiatively efficient and it is possible to estimate the accretion rate from the bolometric luminosity. In the hard state, the accretion flows are thought to be radiatively inefficient -- thus the bolometric luminosity cannot be used directly as a measure of the accretion rate. However, in the hard state one always observes a radio jet. It has been suggested that the jet power is linearly coupled with the accretion rate (see e.g. \cite{FalckeBiermann1995}). This implies that we can use the jet as a tracer of the accretion rate for inefficiently accreting objects. An accretion measure based on the jet has not only the advantage of being usable for the hard state, but is also very easy to use as it only requires a radio core flux.

\section{Accretion rates}
In order to create an estimator of the accretion rate using the radio luminosities, we first have to obtain a sample of steady jet-emitting sources with a measured accretion rate. 

Atoll NSs seem to have steady radio emitting jets \cite{MigliariFender2005b}. As they have a stellar surface, all accreted power has to be radiated away. To compare NSs with BHs, we assume that the boundary layer contributes 1/2 of the total bolometric luminosity \cite{FrankKingRaine2002}. This assumption is not crucial, see \cite{KoerdingFenderMigliari2006} for details.
BHs only have steady jets in their hard states, ie., when the source is thought to be radiatively inefficient and the bolometric luminosity cannot be used to estimate the accretion rate.
Once a source is in the efficiently radiating soft state the jet is quenched. 

\begin{figure}
\resizebox{6cm}{!}{\includegraphics{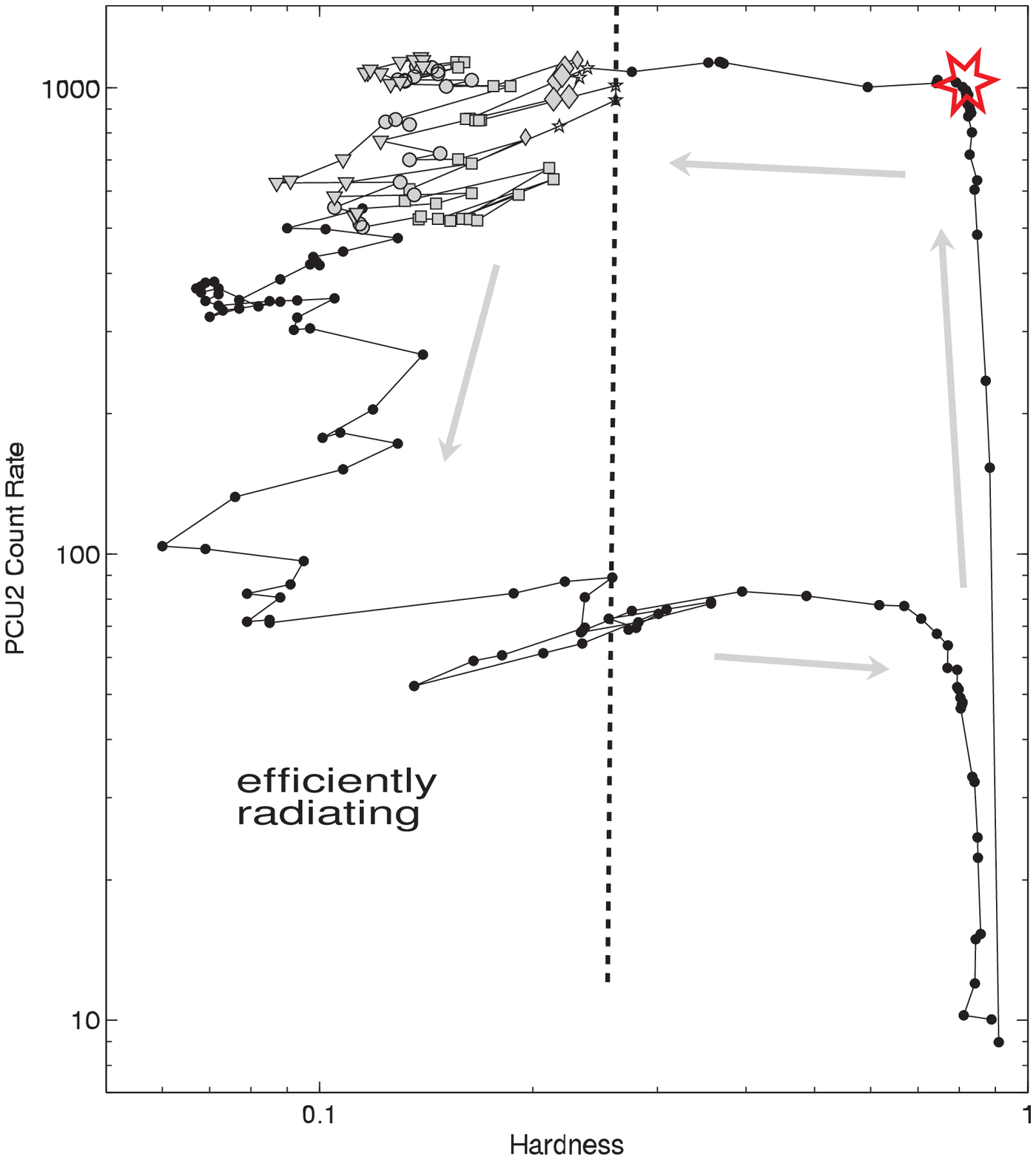}}
\resizebox{8.7cm}{!}{\includegraphics{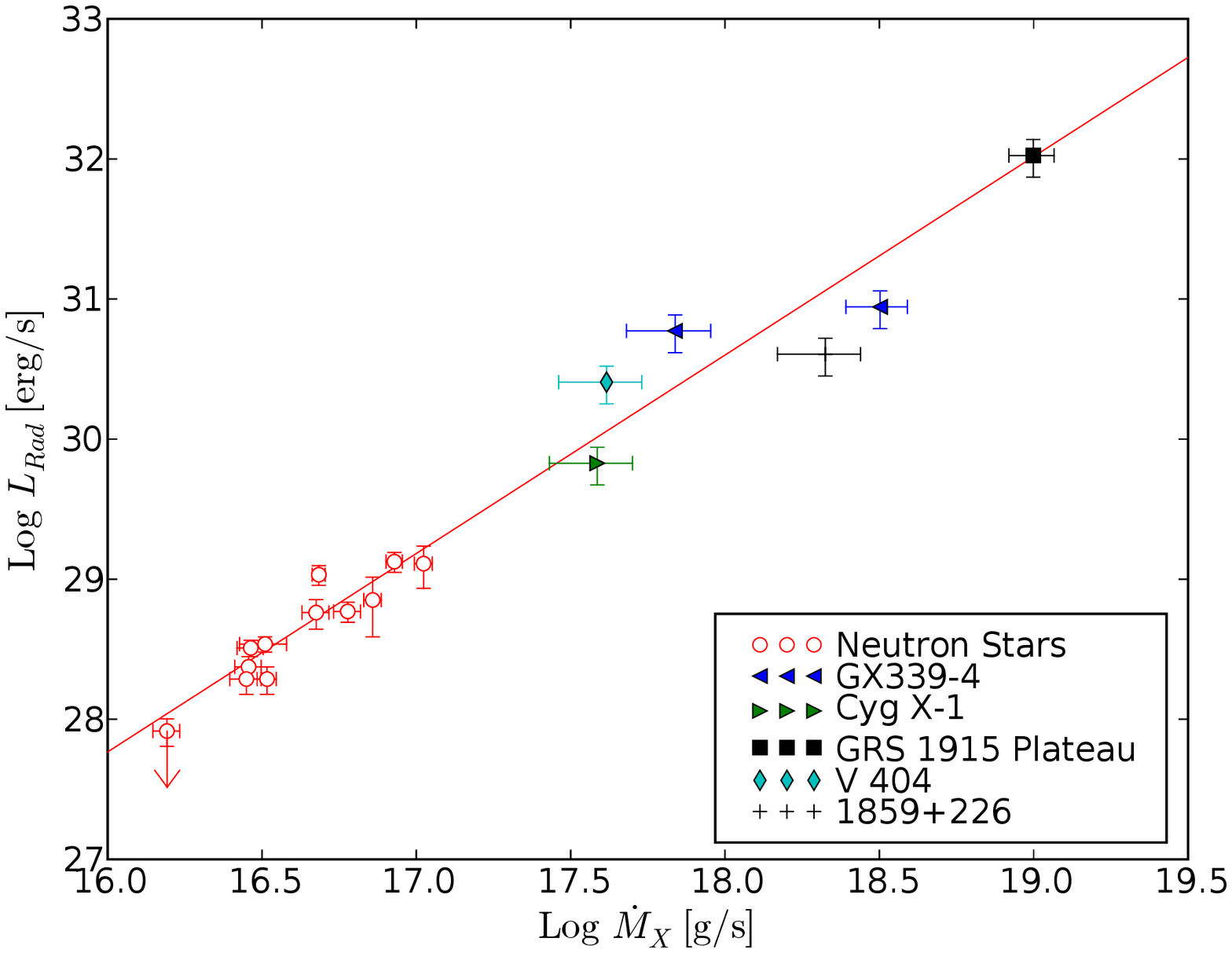}}
\caption{Left side: hardness intensity diagram for GX339-4 (reproduced with permission from Belloni et al 2005 \protect\cite{BelloniHomanCasella2005}). The source is in the hard inefficiently accreting state on the right side of the diagram. As we argue in the text we can only use the datapoint marked with the big red star to normalize our accretion rate measure. At this point the accretion flow is nearly efficiently radiating and still has a steady radio jet. Right side: Radio luminosity as a function of power liberated in the accretion disk for NSs and BHs.}
\label{fiAcc}
\end{figure}

To construct a sample of BHs showing a steady radio jet with estimated accretion rates we can use the fact the the bolometric luminosity does not change significantly during the state transition from the hard to the soft state \cite{ZhangCuiHarmon1997}. The hardness intensity diagram (HID) of the 2003/2004 outburst of GX339-4 is shown on the left side of Fig.~\ref{fiAcc}. While the source is on the left/soft side of the diagram, we can use the bolometric luminosity to estimate the accretion rate. On the right/hard side the source is likely to be radiatively inefficient. The conversion factor to bolometric luminosities for the plotted PCU counts is roughly similar for both the soft and the hard state. 
As the transition in the HID is roughly horizontal, this supports the aforementioned result that the bolometric luminosity does not change during the state transition. If we assume that the accretion rate does not change rapidly during the state transition, we can use the bolometric luminosity just before the state transition to estimate the accretion rate for the source (which still has a steady jet). For every outburst we only obtain a single measurement, which is indicated for the 2003/2004 outburst with the big red star. We found data for 5 outbursts ($2\times$ GX~339-4, Cyg X-1, 1859+226, V404 Cyg). Additionally we included GRS~1915+105, which is constantly accreting near its Eddington limit. In its plateau state it has a steady jet and is likely to be radiatively efficient. 
 
On the right side of Fig.~\ref{fiAcc} we plot the 5 GHz Radio luminosity ($\nu L_\nu$) as a function of the accretion rate for the sample of 5 outbursts and the NSs. To obtain the accretion rate from the bolometric luminosity we assume a constant efficiency of $\eta \approx 0.1$ for both BHs and NSs. For the NSs we have further reduced the bolometric luminosity by a factor 2 to subtract the boundary layer. Fig.~\ref{fiAcc} is therefore a plot of the radio luminosity against power liberated in the accretion disk. We find a strong correlation between radio luminosity and accretion power. The solid line represents the analytical prediction of a conical jet model $L_R \propto \dot{M}^{1.4}$ (e.g., \cite{BlandfordKonigl1979}), which describes the data well.

Thus, we can use the radio luminosity to estimate the accretion rate:
\begin{equation}
\dot{M} = \dot{M}_{\mathrm 0} \left(\frac{L_{\mathrm 8 GHz}}{L_{0,\mathrm 8 GHz}} \right)^{0.71}, \label{eqMdotrad}
\end{equation}
where the normalisation factors ($\dot{M}_{\mathrm 0}$, $L_{0, \mathrm 8GHz}$) is determined by the normalization of the radio / accretion rate correlation. As both constants are exchangeable we set $L_{0, \mathrm 8 GHz} = 10^{30}$ erg s$^{-1}$. This is roughly the 8.6 GHz radio luminosity where the accretion disc around a $10 M_\odot$ BH changes its spectral state. For $\dot{M}_{\mathrm 0}$ we find:
\begin{equation}
\dot{M}_0 = 4.0 \times 10^{17} \mbox{g s}^{-1}.
\end{equation}
If one treats NSs and BHs separately and does not consider the contribution of the boundary layer, the normalizations for both classes differ by roughly a factor 2 (see \cite{KoerdingFenderMigliari2006}).

\begin{figure}
\resizebox{7.5cm}{!}{\includegraphics{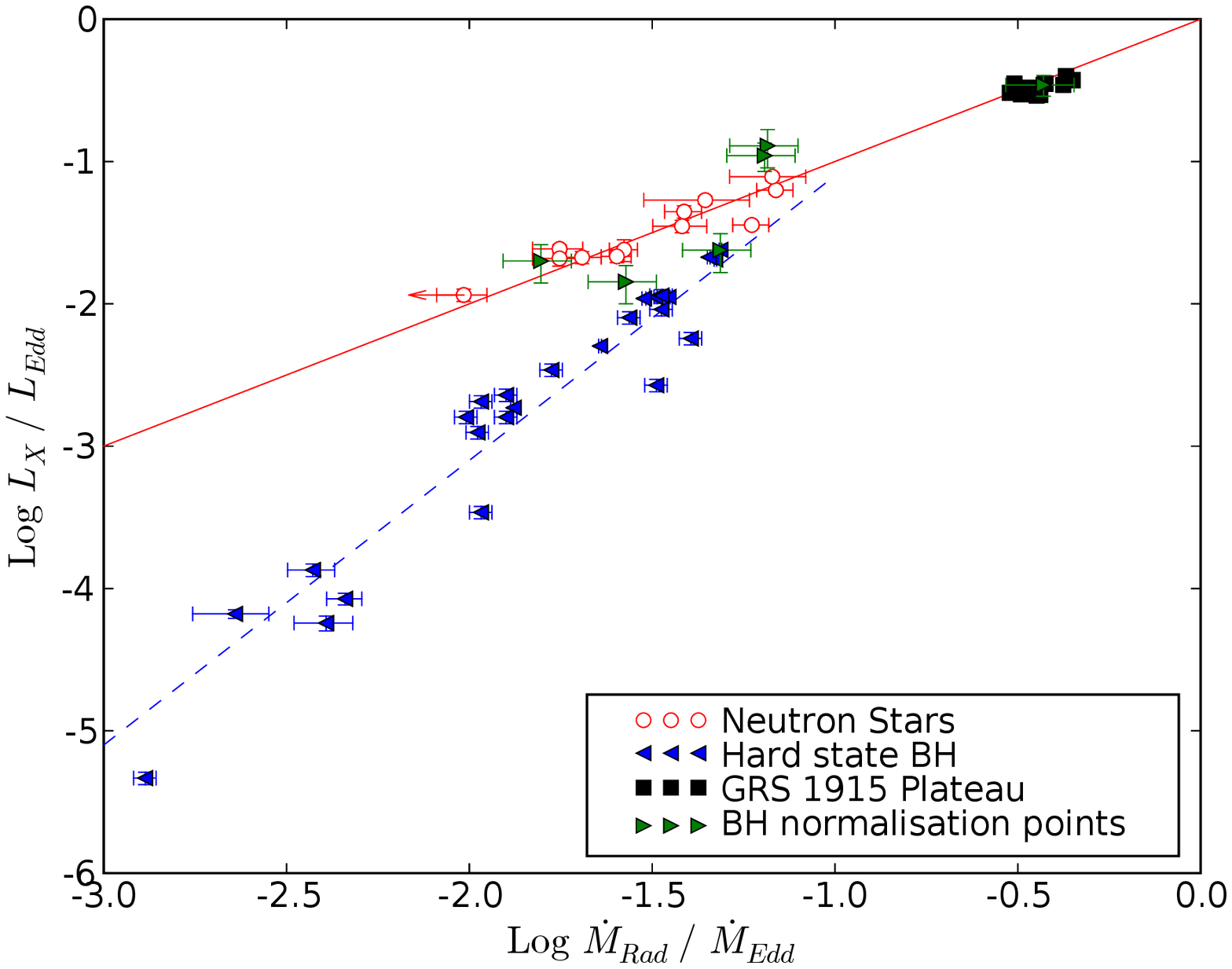}}
\resizebox{7.5cm}{!}{\includegraphics{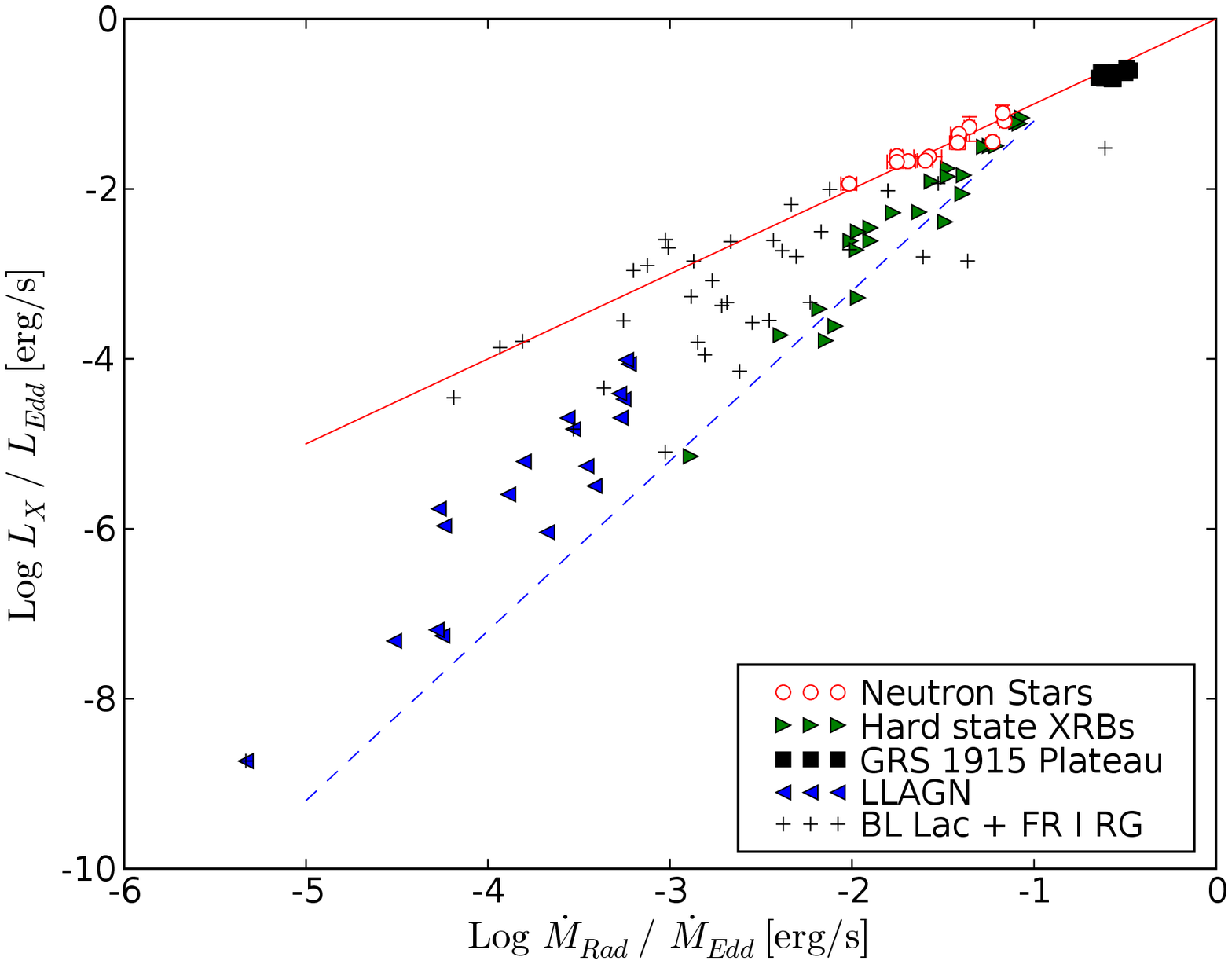}}
\caption{Bolometric luminosity as a function of the accretion rate estimated from the radio luminosity. Left side: Our sample of stellar BHs and NSs. While the bolometric luminosity in NSs depends linearly on accretion rate it scales quadratically for hard state BHs. Right side: Same as left only for AGN assuming the same bolometric correction for AGN as measured in BHs.}
\label{fiNSs}
\end{figure}

This accretion rate estimator was normalized using efficiently radiating sources only. We did not use the observations for which the accretion rate was to be estimated for the normalization of our method.
 Thus, we can compare the bolometric luminosity with our accretion rate estimations. On the left side of Fig.~\ref{fiNSs} we plot the bolometric X-ray luminosity against accretion rate for BH and NS XRBs. The NSs and the efficiently accreting source GRS~1915+105 have bolometric luminosities $L_{bol} \propto \dot{M}$. However, this only reflects that the radio luminosity follows the predictions of a jet model. The hard state black hole significantly deviate from the linear scaling found for NSs. We find that they scale like $L_{bol} \propto \dot{M}^2$. We note that this quadratic dependence has not been put into the accretion measure, only the linear dependence of the NSs and efficient accreting objects has been used to normalize the accretion measure.

At low accretion rates of 0.001 $\dot{M}_{Edd}$ BHs have significantly lower bolometric luminosities than NSs ($\sim$ two orders of magnitude). We can use this difference to explore if BHs advect some of the accreted power. All accreted power has to leave the accretion flow in some way. Energy can escape from the accreting system only in the form of matter (wind and jet) and as radiation.
\begin{equation}
\dot{M} \eta c^2 = P_{jet} + P_{wind} + L_{bol} + P_{Advect},
\end{equation}
where $\eta$ denotes the efficiency with that the BH can create energy from accretion (e.g., $\eta \approx 0.1$), $L_{bol}$ denotes the bolometric luminosity, $P_{jet}$ and $P_{wind}$ the power injected into the jet and wind and $P_{Advect}$ is the power advected into the BH. The latter term vanishes for a NS as advection is impossible due to the stellar surface. We attribute all matter and magnetic fields not contained in the jet to the accretion disc wind. 

Assuming that the radiative efficiency of BH and NS are similar, we have seen that
\begin{equation}
P_{jet}^{NS}(\dot{M}) \approx P_{jet}^{BH}(\dot{M}) \approx q_j \dot{M},
\end{equation}
Disk winds play an important role for strongly accreting sources near the Eddington limit. For strongly sub-Eddington sources winds are likely to be less important for the energy budget. But even if winds carry a significant fraction of the accreted power, it is likely that NSs and BHs have similar wind properties due to their similarities in the SED and the power spectral density. We note that NSs have larger bolometric luminosity, thus if the wind properties of both classes are different NSs are probably the ones with the stronger radiation driven wind. In equations:
\begin{equation}
P_{wind}^{NS}(\dot{M})\approx P_{wind}^{BH}(\dot{M}).
\end{equation}
Combining these formulae, we find:
\begin{equation}
P_{Advect}^{BH} \approx L^{NS}_{bol}(\dot{M}) - L^{BH}_{bol}(\dot{M})
\end{equation}
For strongly sub-Eddington sources this difference is larger than the total bolometric luminosity. Those sources are therefore likely to advect more energy than they radiate in X rays.

\section{Active Galactic Nuclei}

\begin{figure}
\resizebox{7.5cm}{!}{\includegraphics{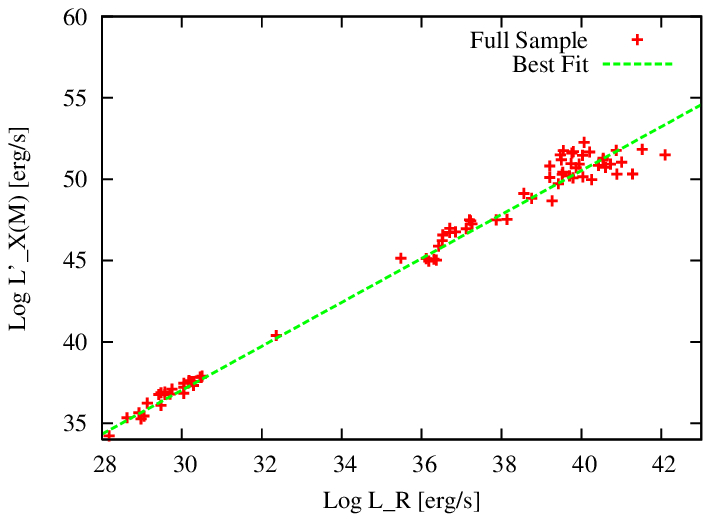}}
\resizebox{7.5cm}{!}{\includegraphics{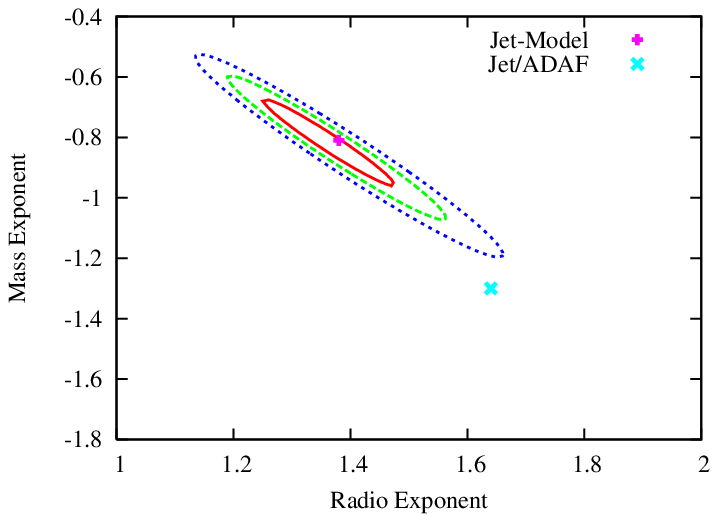}}
\caption{Fundamental plane of accreting black holes in the version of \protect\cite{KoerdingFalckeCorbel2005}. Left side: X-ray luminosity scaled for BH mass against radio luminosity for a sample of XRBs, LLAGN, FR-I RGs and BL Lac objects. Right side: $\chi^2$ map for the mass and radio luminosity correlation parameter. One, two and three sigma contours are shown.}
\label{fifun}
\end{figure}

The fundamental plane of accreting black holes \cite{MerloniHeinzdiMatteo2003,FalckeKoerdingMarkoff2004,KoerdingFalckeCorbel2005} connects XRBs and AGN with a plane in the BH mass, X-ray and radio luminosity space. The fundamental plane can be parametrized as 
\begin{equation}
\log L_X = \left( 1.41 \pm 0.11 \right) \log L_{Rad} - \left(0.87 \pm 0.14\right) \log M + b,
\end{equation}
where $b$ is the unimportant offset. One projection of the plane is shown on the left side of Fig.~\ref{fifun}. The right side of the figure presents the corresponding $\chi^2$ map. We note that the significance contours are highly elliptical. So it is possible to increase the radio coefficient a little while reducing the mass coefficient without changing the $\chi^2$ significantly, e.g., the parameter pair (1.5, -1) is just barely out of the $1 \sigma$ contour while (1.4, -1) can be rejected with more than 3 $\sigma$. 

The exact parameters depend slightly on the sample used to derive the fundamental plane. If one puts different weights on the different classes of AGN (e.g., number of LLAGN compared to FR-I RGs or Quasars) one finds slightly different parameters \cite{KoerdingFalckeCorbel2005}. We have shown in \cite{MerloniKoerdingHeinz2006} that the Fundamental plane is not affected by huge differences in the distance of XRBs and AGN. The correlation is still visible if one plots only fluxes and has been checked using partial correlation analysis and Monte Carlo simulations.

Using our accretion rate measure, we can rewrite the fundamental plane to:
\begin{equation}
\frac{L_X}{L_{Edd}} \propto \left(\frac{\dot{M}}{\dot{M}_{Edd}}\right)^2 M^{0.14} . \label{eqFundamental}
\end{equation}
This is, up to the factor $M^{0.14}$, exactly what one would expect for an inefficient accretion flow around a black hole. This factor $M^{0.14}$ can be seen as a consequence of our choice of the dependence of the accretion rate on radio luminosity ($L_R\propto \dot{M}^{1.4}$). If we had used a larger value (e.g., 1.5), which would still be in agreement with the data, the factor would nearly vanish. In that case the exponent for the radiatively inefficient accretion is not exactly 2 as we currently find. Additionally we note that the fundamental plane has been derived using 2-10 keV X-ray luminosities, while we used bolometric luminosities in our studies here. The correction factor $M^{0.14}$ may therefore be real, ie. a bolometric correction varying with black hole mass or it can represent the microphysics that is not scale invariant. 

\begin{figure}
\resizebox{7.5cm}{!}{\includegraphics{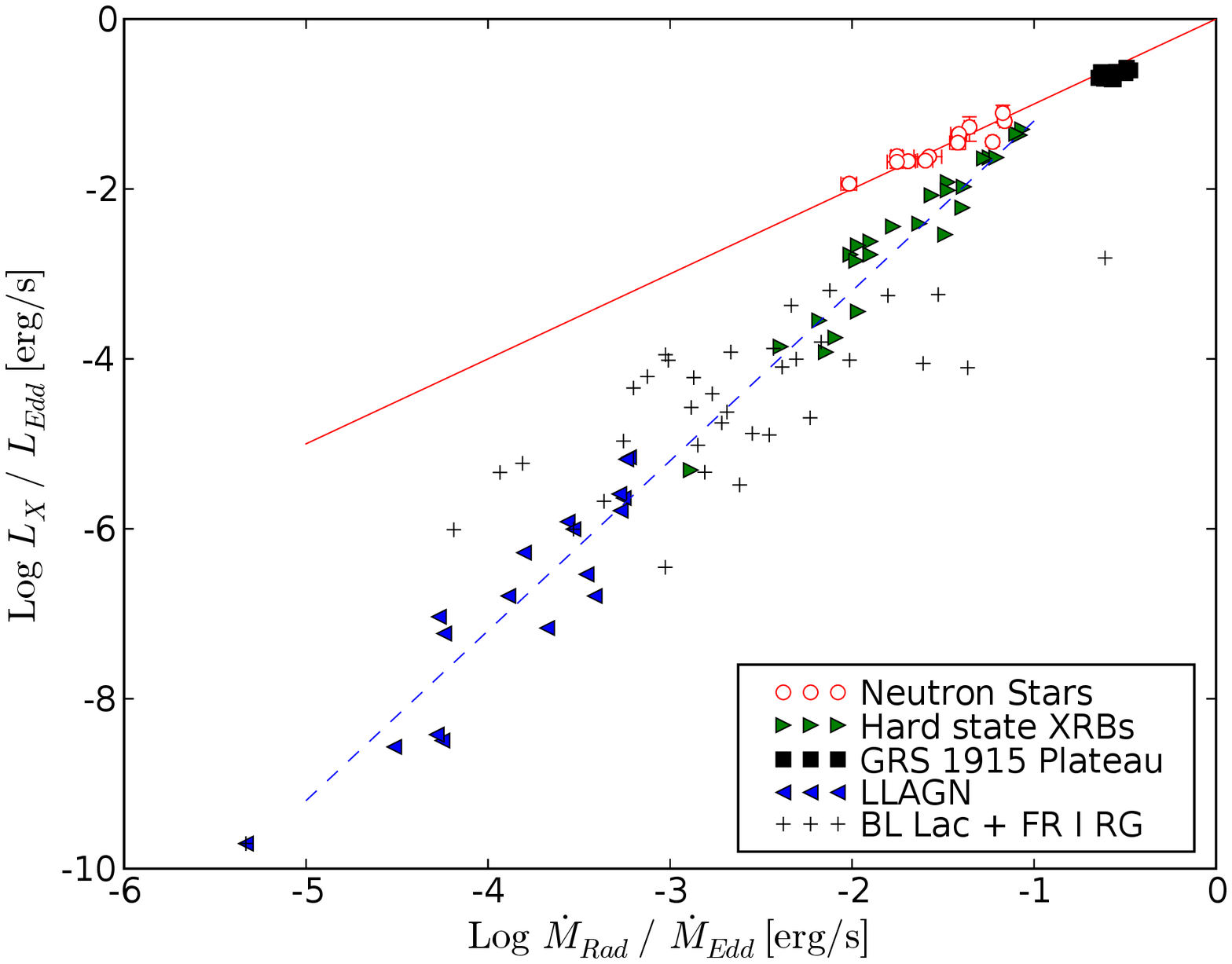}}
\resizebox{7.5cm}{!}{\includegraphics{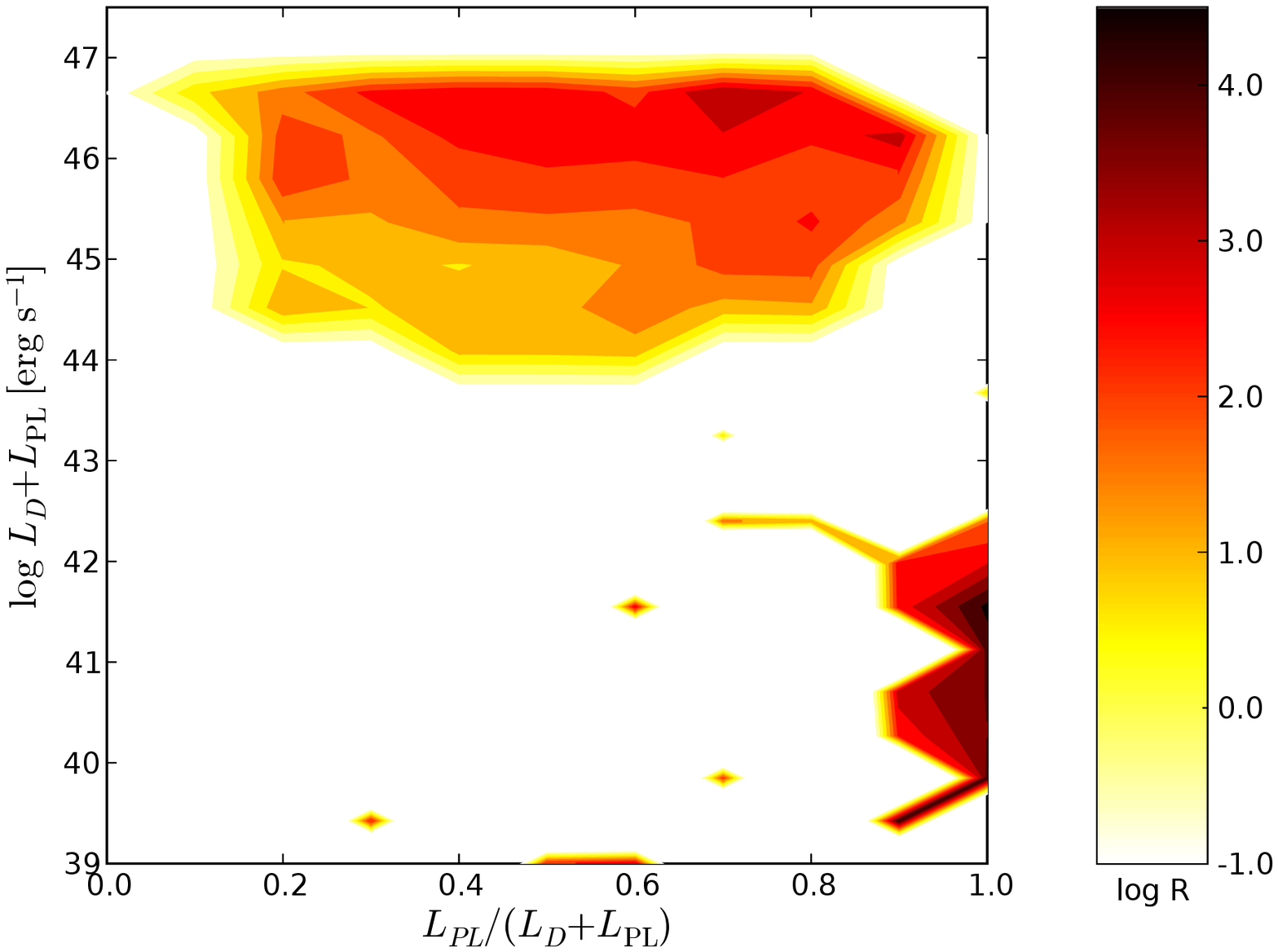}}
\caption{Right side: Bolometric luminosity as a function of accretion rate. In comparison to Fig. 2 we now include the correction factor $M^{0.14}$ originating from the fundamental plane. The scatter is significantly reduced. Left side: Disk-fraction luminosity diagram of a sample of AGN. It is a generalization of HIDs used in XRBs, see Fig. 1 for a HID of GX 339-4. AGN and XRBs populate the same regions in the diagram. The fundamental plane and our accretion measure is only valid on the right side of the diagram.}
\label{fiMap}
\end{figure}

If we use the same bolometric correction for XRBs as for AGN, we find the plot shown on the right side of Fig.~\ref{fiNSs}. Similar to the stellar BHs LLAGN are already significantly below the linear correlation expected for efficiently radiating sources. The large scatter around the inefficient scaling is due to the missing mass correction factor $M^{0.14}$. If we include that factor as "effective bolometric correction" (whether it is a real effect or a relic of our choice of the accretion measure), we arrive at Fig.~\ref{fiMap} (left panel). Now the inefficient scaling found in stellar BHs is also found in hard state AGN: LLAGN, FR-I RGs and BL Lac objects.

One of the differences between the two suggested versions of the fundamental plan is that Falcke and K\"ording et al.   \cite{FalckeKoerdingMarkoff2004,KoerdingFalckeCorbel2005} try to include only hard-state objects, while Merloni et al. \cite{MerloniHeinzdiMatteo2003} try to include a representative sample of all AGN, including quasars which probably belong to the soft state. Both studies obtain similar results for the fundamental plane. Here, we explore why the inclusion of soft state objects seem to effect only the scatter around the correlation.
First, we note that if a source is brighter than the luminosity where the inefficient scaling reaches the track of efficient accretion, it cannot continue to scale as $\dot{M}^2$, but its bolometric luminosity needs to scale linearly with accretion rate. This suggests, that high accretion rate AGN should have a reduced X-ray flux compared to the correlation (up to a factor $\sim 30 - 100$). 
 
In the picture presented in \cite{KoerdingJesterFender2006} the fundamental plane is only valid on the right side of an disk-fraction luminosity diagram (DFLD) (see Fig.~\ref{fiMap} right panel and the contribution by Jester et al. these proceedings). For low accretion rates (below 1\% Edd) all objects are on the right side of the diagram and show strong radio jets. For larger accretion rates the sources can either be radio loud or quiet, the probablity to be in on of those states seems to depend on their position in the DFLD. If we include a sample of radio quiet and loud AGN, the average radio luminosity will be reduced compared to what one would expect from our accretion rate measure (see e.g., \cite{MaccaroneGalloFender2003}). This compensates the previously mentioned reduction of the X-ray flux due to the end of the quadratic scaling. Inclusion of radio loud and quite AGN will therefore mainly increase the scatter around the correlation.


\begin{thebibliography}{10}

\bibitem{BelloniHomanCasella2005}
T.~{Belloni}, J.~{Homan}, P.~{Casella}, M.~{van der Klis}, E.~{Nespoli},
  W.~H.~G. {Lewin}, J.~M. {Miller}, and M.~{M{\'e}ndez}, \emph{{The evolution
  of the timing properties of the black-hole transient GX 339-4 during its
  2002/2003 outburst}}, \aap\, \textbf{440}, 207--222 (2005).

\bibitem{BlandfordKonigl1979}
R.~D. {Blandford} and A.~{K\"onigl}, \emph{Relativistic jets as compact radio
  sources}, \apj\, \textbf{232}, 34--48 (1979).

\bibitem{CorbelFenderTzioumis2000}
S.~{Corbel}, R.~P. {Fender}, A.~K. {Tzioumis}, M.~{Nowak}, V.~{McIntyre},
  P.~{Durouchoux}, and R.~{Sood}, \emph{{Coupling of the X-ray and radio
  emission in the black hole candidate and compact jet source GX 339-4}},
  \aap\, \textbf{359}, 251--268 (2000).

\bibitem{EsinMcClintockNarayan1997}
A.~A. {Esin}, J.~E. {McClintock}, and R.~{Narayan}, \emph{{Advection-dominated
  Accretion and the Spectral States of Black Hole X-Ray Binaries: Application
  to Nova MUSCAE 1991}}, \apj\, \textbf{489}, 865+ (1997).

\bibitem{FalckeBiermann1995}
H.~{Falcke} and P.~L. {Biermann}, \emph{{The jet-disk symbiosis. I. Radio to
  X-ray emission models for quasars.}}, \aap\, \textbf{293}, 665--682 (1995).

\bibitem{FalckeKoerdingMarkoff2004}
H.~{Falcke}, E.~{K\"ording}, and S.~{Markoff}, \emph{{A scheme to unify
  low-power accreting black holes. Jet-dominated accretion flows and the
  radio/X-ray correlation}}, \aap\, \textbf{414}, 895--903 (2004).

\bibitem{FenderCorbelTzioumis1999}
R.~{Fender}, S.~{Corbel}, T.~{Tzioumis}, V.~{McIntyre}, D.~{Campbell-Wilson},
  M.~{Nowak}, R.~{Sood}, R.~{Hunstead}, A.~{Harmon}, P.~{Durouchoux}, and
  W.~{Heindl}, \emph{{Quenching of the Radio Jet during the X-Ray High State of
  GX 339-4}}, \apjl\, \textbf{519}, L165--L168 (1999).

\bibitem{Fender2001}
R.~P. {Fender}, \emph{{Powerful jets from black hole X-ray binaries in low/hard
  X-ray states}}, \mnras\, \textbf{322}, 31--42 (2001).

\bibitem{FrankKingRaine2002}
J.~{Frank}, A.~{King}, and D.~J. {Raine}, \emph{{Accretion Power in
  Astrophysics: Third Edition}}, Accretion Power in Astrophysics, by Juhan
  Frank and Andrew King and Derek Raine, pp.~398.~ISBN 0521620538.~Cambridge,
  UK: Cambridge University Press, February 2002., February 2002.

\bibitem{GierlinskiDone2004}
M.~{Gierli{\'n}ski} and C.~{Done}, \emph{{Black hole accretion discs: reality
  confronts theory}}, \mnras\, \textbf{347}, 885--894 (2004) [{\tt
  astro-ph/0307333}].

\bibitem{KoerdingFalckeCorbel2005}
E.~{K{\"o}rding}, H.~{Falcke}, and S.~{Corbel}, \emph{{Refining the fundamental
  plane of accreting black holes}}, \aap\, \textbf{456}, 439--450 (2006) [{\tt
  astro-ph/0603117}].

\bibitem{KoerdingFenderMigliari2006}
E.~G. {K{\"o}rding}, R.~P. {Fender}, and S.~{Migliari}, \emph{{Jet-dominated
  advective systems: radio and X-ray luminosity dependence on the accretion
  rate}}, \mnras\, \textbf{369}, 1451--1458 (2006).

\bibitem{KoerdingJesterFender2006}
E.~G. {K{\"o}rding}, S.~{Jester}, and R.~{Fender}, \emph{{Accretion states and
  radio loudness in active galactic nuclei: analogies with X-ray binaries}},
  \mnras\, \textbf{372}, 1366--1378 (2006) [{\tt astro-ph/0608628}].

\bibitem{MaccaroneGalloFender2003}
T.~J. {Maccarone}, E.~{Gallo}, and R.~{Fender}, \emph{{The connection between
  radio-quiet active galactic nuclei and the high/soft state of X-ray
  binaries}}, \mnras\, \textbf{345}, L19--L24 (2003).

\bibitem{MarkoffFalckeFender2001}
S.~{Markoff}, H.~{Falcke}, and R.~{Fender}, \emph{{A jet model for the
  broadband spectrum of XTE J1118+480. Synchrotron emission from radio to
  X-rays in the Low/Hard spectral state}}, \aap\, \textbf{372}, L25--L28
  (2001).

\bibitem{MarkoffNowakWilms2005}
S.~{Markoff}, M.~A. {Nowak}, and J.~{Wilms}, \emph{{Going with the Flow: Can
  the Base of Jets Subsume the Role of Compact Accretion Disk Coronae?}},
  \apj\, \textbf{635}, 1203--1216 (2005) [{\tt astro-ph/0509028}].

\bibitem{MerloniHeinzdiMatteo2003}
A.~{Merloni}, S.~{Heinz}, and T.~{Di Matteo}, \emph{{A Fundamental Plane of
  black hole activity}}, \mnras\, \textbf{345}, 1057--1076 (2003).

\bibitem{MerloniKoerdingHeinz2006}
A.~{Merloni}, E.~{K{\"o}rding}, S.~{Heinz}, S.~{Markoff}, T.~{Di Matteo}, and
  H.~{Falcke}, \emph{{Why the fundamental plane of black hole activity is not
  simply a distance driven artifact}}, New Astronomy\, \textbf{11}, 567--576
  (2006) [{\tt astro-ph/0601286}].

\bibitem{MigliariFender2005b}
S.~{Migliari} and R.~P. {Fender}, \emph{{Jets in neutron star X-ray binaries: a
  comparison with black holes}}, \mnras\, \textbf{366}, 79--91 (2006) [{\tt
  astro-ph/0510698}].

\bibitem{NarayanYi1994}
R.~{Narayan} and I.~{Yi}, \emph{{Advection-dominated accretion: A self-similar
  solution}}, \apjl\, \textbf{428}, L13--L16 (1994).

\bibitem{Nowak1995}
M.~A. {Nowak}, \emph{{Toward a Unified View of Black-Hole High-Energy States}},
  \pasp\, \textbf{107}, 1207+ (1995).

\bibitem{ShakuraSunyaev1973}
N.~I. {Shakura} and R.~A. {Sunyaev}, \emph{Black holes in binary systems.
  observational appearance.}, \aap\, \textbf{24}, 337--355 (1973).

\bibitem{SunyaevTruemper1979}
R.~A. {Sunyaev} and J.~{Tr\"umper}, \emph{{Hard X-ray spectrum of CYG X-1}},
  \nat\, \textbf{279}, 506--508 (1979).

\bibitem{ThornePrice1975}
K.~S. {Thorne} and R.~H. {Price}, \emph{{Cygnus X-1 - an interpretation of the
  spectrum and its variability}}, \apjl\, \textbf{195}, L101--L105 (1975).

\bibitem{ZhangCuiHarmon1997}
S.~N. {Zhang}, W.~{Cui}, B.~A. {Harmon}, W.~S. {Paciesas}, R.~E. {Remillard},
  and J.~{van Paradijs}, \emph{{The 1996 Soft State Transition of Cygnus X-1}},
  \apjl\, \textbf{477}, L95 (1997).

\end{thebibliography}
\end{document}